\documentclass[conference]{IEEEtran}
\IEEEoverridecommandlockouts
\usepackage{cite}
\usepackage{amsmath,amssymb,amsfonts}
\usepackage{algorithmic}
\usepackage{graphicx}
\usepackage{textcomp}
\usepackage{xcolor}
\usepackage[hyperfootnotes=false]{hyperref}

\usepackage{tikz}
\usetikzlibrary{shapes.geometric, arrows, positioning, shadows}
\usepackage{pgf-pie}
\usepackage{pgfplots}
\pgfplotsset{compat=1.17}

\usepackage{enumitem}

\usepackage{tabularx}

\def\BibTeX{{\rm B\kern-.05em{\sc i\kern-.025em b}\kern-.08em
    T\kern-.1667em\lower.7ex\hbox{E}\kern-.125emX}}
\begin{document}

\title{From Textual to Verbal Communication: \\Towards Applying Sentiment Analysis \\to a Software Project Meeting
}

\author{\IEEEauthorblockN{Marc Herrmann}
\IEEEauthorblockA{Leibniz University Hannover\\
Software Engineering Group\\
Hannover, Germany\\
marc.herrmann@stud.uni-hannover.de}
\and
\IEEEauthorblockN{Jil Klünder}
\IEEEauthorblockA{Leibniz University Hannover\\
Software Engineering Group\\
Hannover, Germany \\
jil.kluender@inf.uni-hannover.de}
}

\maketitle

\begin{abstract}
Sentiment analysis gets increasing attention in software engineering with new tools emerging from new insights provided by researchers. Existing use cases and tools are meant to be used for textual communication such as comments on collaborative version control systems. While this can already provide useful feedback for development teams, a lot of communication takes place in meetings and is not suited for present tool designs and concepts. 

In this paper, we present a concept that is capable of processing live meeting audio and classifying transcribed statements into sentiment polarity classes. We combine the latest advances in open source speech recognition with previous research in sentiment analysis. We tested our approach on a student software project meeting to gain proof of concept, showing moderate agreement between the classifications of our tool and a human observer on the meeting audio. Despite the preliminary character of our study, we see promising results motivating future research in sentiment analysis on meetings. For example, the polarity classification can be extended to detect destructive behaviour that can endanger project success.   
\end{abstract}


\section{Introduction}
Given the increasing complexity and size of software projects, software development is a team effort rather than a one-person activity~\cite{kraut1995coordination}. Working in a team requires social interactions and adequate communication, as the success of a software project depends on the quality of the collaboration. Research has shown that happy developers are more productive and solve problems better than dissatisfied ones~\cite{graziotin2014happy,graziotin2015you}. Hence, project leaders and managers are interested in being aware of the temporary emotional shade in the team, which we refer to as mood. One method to gain an overview of the team mood on-the-fly is known as \textit{sentiment analysis} which analyzes communication with respect to the transported polarity or other sentiments~\cite{bakshi2016opinion}. So far, sentiment analysis has been frequently applied in software engineering, as a recent systematic literature study shows~\cite{obaidi2021development}. Sentiment analysis is applied to a wide variety of data sources, including JIRA, GitHub, and Stack Overflow~\cite{obaidi2021development}. However, all these data sources provide textual communication such as text-messages, comments, tickets, and the like. 

Despite the increasing amount of decentralized software development~\cite{kuhrmann2018helena}, a lot of communication takes part in meetings. Based on the results of a fine-grained interaction analysis applied to 32 student software project meetings, Schneider et al.~\cite{schneider2018positive} show the relevance of specific types of statements enabling to increase the positive team mood after the meeting. There are other attempts to analyze interactions in meetings, including a coding scheme adjusted for software projects~\cite{prenner2018making,klunder2020you}, but these methods still require manual effort leading to subjective results. 

In a first attempt to allow for meeting analysis in real-time providing objective results, we want to \textit{apply sentiment analysis to verbal communication in meetings}. In this paper, as a first step, we present an approach that processes verbal communication to prepare a transcript that can be used as input for different sentiment analysis tools. We base our work on the sentiment analysis tool presented by Klünder et al.~\cite{kluender2020identifying} and adjust it to be suitable for verbal communication. The preliminary application of the tool, which we refer to as the \textit{SEnti-Analyzer}, to a student software project meeting provides two promising results:

\begin{itemize}
    \item[(1)] Sentiment analysis can be applied to meetings, and
    \item[(2)] The application of sentiment analysis to verbal communication is as meaningful as the application to textual communication.
\end{itemize}

Furthermore, our exemplary application in the case study reveals interesting and relevant aspects of future work.

\textit{Context.} This paper is based on Herrmann's bachelor thesis~\cite{herrmann2021automatic} entitled ``Automatic Classification of Statements in Meetings of Development Teams''. 

\textit{Outline.} The rest of the paper is structured as follows: In Section~\ref{sec:background}, we highlight related research and background details. Section~\ref{sec:research} introduces the concept and its application in the case study. The results are presented in Section~\ref{sec:results} and interpreted in Section~\ref{sec:interpretation}. The paper is summarized in Section~\ref{sec:conclusions}.

\section{Background and Related Work}
\label{sec:background}
Meeting analysis and sentiment analysis have both been frequently applied to software projects. For example, Klünder et al.~\cite{klunder2020you} elaborate the coding scheme \textit{act4teams-SHORT}, which they derived from an established interaction analysis scheme in psychology. Using \textit{act4teams-SHORT}, statements in a meeting can be categorized in one of eleven different categories such as ``naming problems'' or ``giving information''. According to the results of Klünder et al. \cite{klunder2020you}, using this coding scheme and analyzing the resulting interactions in each category help identifying possible problematic behavior. Resolving this kind of behavior at early stages of a software project can lead to better overall team performance and project success \cite{klunder2020you}. This categorization of high level interaction analysis can be traced back to basic low level sentiment analysis which finds more and more applications recently.

There are numerous different sentiment analysis tools available, even some especially related to software engineering~\cite{calefato2018senti}. However, tools for languages differing from English are still rare. Klünder et al.~\cite{kluender2020identifying} developed a classifier for German text messages from group chats of development teams that maps the input data to the polarity of the message, i.e., \textit{positive}, \textit{negative}, or \textit{neutral}. This classifier is based on a trained classification model and defines a key part of the \textit{SEnti-Analyzer} which we present in this paper. Calefato et al.~\cite{calefato2018senti} present their tool \textit{Senti4SD} to provide a sentiment analysis tool trained on the software engineering domain (using data from \textit{Stack Overflow}), which does not result in the misclassification of the associated terminology. A similar approach is used by Islam et al.~\cite{islam2017leveraging}, using \textit{JIRA} issue comments for training their tool named \textit{SentiStrength-SE}, and Ahmed et al.~\cite{ahmed2017senticr}, who use code review comments for training their tool \textit{SentiCR}.

Besides sentiment analysis, our approach is related to previous research in speech recognition. 
Agarwal and Zesch~\cite{agarwalzesch2019german} present their approach in training a German-language model for the \textit{Mozilla DeepSpeech} framework, which also constitutes the foundation of our speech recognition. The framework provides a transcript that can be used as input for existing sentiment analysis tools. 

Speech recognition has also been applied to meetings in software engineering with another focus:
Gall and Berenbach~\cite{gall2006towards} present a framework recording requirements elicitation meetings on video, thereby collecting relevant information raised by stakeholders. Shakeri et al.~\cite{abad2018elica} also strive to extract relevant information presented in elicitation meetings. Their tool \textit{ELICA} collects knowledge and information related to requirements. This way, it helps analyzing the meeting outcome. 

In this paper, we combine the approaches of sentiment analysis tools with automatic speech recognition to analyze verbal team communication in real-time. Our tool provides an overview of the distribution of sentiment categories for the recognized statements at the end of the meeting. This way the project manager can easily gain first direct feedback about the course of the meeting.

Although both sentiment analysis and meeting analysis have been proven to be beneficial for software projects, to the best of our knowledge sentiment analysis has not yet been used for meeting analysis. 

\section{Study Design}
\label{sec:research}
In the following, we present our research objective, the research questions, and the study. Our approach basically consists of two steps: (1) the transcription of a meeting and (2) the application of a sentiment analysis tool~\cite{kluender2020identifying}.  

\subsection{Research Objective and Research Questions}
The main objective of our research is to \textit{analyze the sentiments transported in statements made in a meeting of a software project}. To reach this goal, we developed and evaluated a concept and a corresponding software tool, the so-called \textit{SEnti-Analyzer}, which uses an audio stream of verbal (meeting) communication as input and predicts the polarity of each statement. We formulated the following research questions:

\begin{itemize}[leftmargin=.38in]
    \item[RQ1:] How can automatic speech recognition and sentiment analysis be combined to analyze the statements in meetings of a software project? 
    \item[RQ2:] How do the automatically produced results differ from the subjective analysis of a human observer?
\end{itemize}

\subsection{Instrument Development}
Our approach for the \textit{SEnti-Analyzer} is to feed the users microphone input into our software, e.g., the microphone of a laptop placed in the middle of a conference table during the meeting. Note that multiple audio inputs (e.g., as in online conferences) are also possible. 

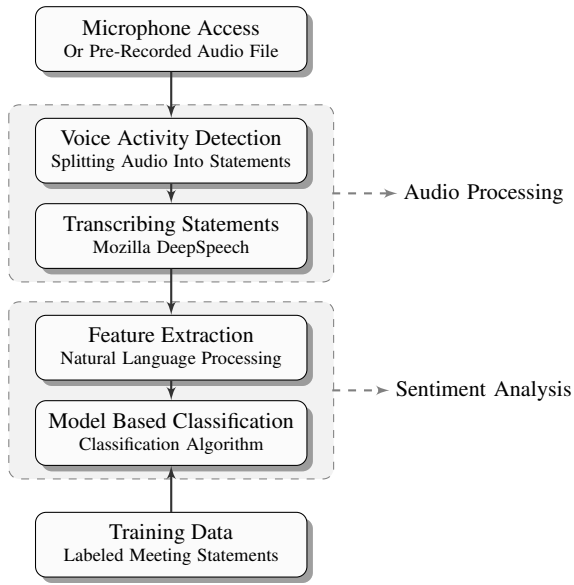
\begin{figure}[htbp]
\centering

\pgfdeclarelayer{background}
\pgfdeclarelayer{foreground}
\pgfsetlayers{background,main,foreground}

\tikzstyle{materia}=[draw, fill=gray!3, text width=6.0em, text centered, minimum height=1.5em,drop shadow]
\tikzstyle{etape} = [materia, text width=14em, minimum width=10em, minimum height=3.5em, rounded corners, drop shadow]
\tikzstyle{texto} = [above, text width=6em, text centered]
\tikzstyle{linepart} = [draw, thick, color=black!50, -latex', dashed]
\tikzstyle{line} = [draw, thick, color=black!80, -latex']
\tikzstyle{ur}=[draw, text centered, minimum height=0.01em]

\newcommand{\blockdist}{1.3}
\newcommand{\edgedist}{1.5}

\newcommand{\etape}[2]{node (p#1) [etape]
  {#2}}

\newcommand{\background}[5]{%
  \begin{pgfonlayer}{background}
    \path (#1.west |- #2.north)+(-0.5,0.25) node (a1) {};
    \path (#3.east |- #4.south)+(+0.5,-0.25) node (a2) {};
    \path[fill=gray!10,rounded corners, draw=black!50, dashed]
      (a1) rectangle (a2);
      \path (#3.east |- #2.north)+(0,0.25)--(#1.west |- #2.north) node[midway] (#5-n) {};
      \path (#3.east |- #2.south)+(0,-0.35)--(#1.west |- #2.south) node[midway] (#5-s) {};
      \path (#3.east |- #2.north)+(0.7,0)--(#3.east |- #4.south) node[midway] (#5-w) {};
  \end{pgfonlayer}}

\newcommand{\transreceptor}[3]{%
  \path [linepart] (#1.east) -- node [above]
    {\normalsize #2} (#3);}

\begin{tikzpicture}[scale=0.7,transform shape]

\path \etape{1}{\large Microphone Access\\\normalsize Or Pre-Recorded Audio File};

\path (p1.south)+(0.0,-1.5) \etape{2}{\large Voice Activity Detection\\\normalsize Splitting Audio Into Statements};
\path (p2.south)+(0.0,-1.0) \etape{3}{\large Transcribing Statements\\\normalsize Mozilla DeepSpeech};

\path (p3.south)+(0.0,-1.5) \etape{4}{\large Feature Extraction\\\normalsize Natural Language Processing};
\path (p4.south)+(0.0,-1.0) \etape{5}{\large Model Based Classification\\\normalsize Classification Algorithm};

\path (p5.south)+(0.0,-1.5) \etape{6}{\large Training Data\\\normalsize Labeled Meeting Statements} (p6);

\path [line] (p1.south) -- node [above] {} (p2);
\path [line] (p2.south) -- node [above] {} (p3);
\path [line] (p3.south) -- node [above] {} (p4);
\path [line] (p4.south) -- node [above] {} (p5);
\path [line] (p6.north) -- (p5.south) {};

\background{p2}{p2}{p3}{p3}{bk1}
\background{p4}{p4}{p5}{p5}{bk2}

\path (bk1-w)+(+3.0,0) node (ur1)[] {\large Audio Processing};
\path (bk2-w)+(+3.0,0) node (ur2)[] {\large Sentiment Analysis};
\transreceptor{bk1-w}{}{ur1};
\transreceptor{bk2-w}{}{ur2};
\end{tikzpicture}

\caption{Simplified processing pipeline of the \textit{SEnti-Analyzer}}
\label{fig:processing}
\end{figure}

The sequence of processing steps from the raw audio input to the resulting prediction of sentiment categories is visualized in Figure~\ref{fig:processing}. After the meeting the user can stop the recording and instantly receives the transcript. This will be generated right during the meeting using the \textit{Mozilla Deepspeech} framework, alongside with the German language models\footnote{Note that the focus on German is due to the nature of the bachelor thesis as outlined in the introduction. Future work will focus on extending the approach to English.}. We use state-of-the-art voice activity detection to separate the audio stream into frames of statements. Stopping the recording also starts the application of natural language processing to the transcript. Once completed, the collected statements and corresponding metrics are fed into the sentiment analysis tool provided by Klünder et al.~\cite{kluender2020identifying} which interprets the results. Finally, an output of classified interactions is presented to the user, e.g., by calculating the total and relative proportions of each category. These steps are fully automated, requiring the user only to start the tool and specifying the end of recording by a single key stroke. The potential for improvement offered by this preliminary tool is outlined in the end of the paper.

\subsection{The Case Meeting}
\label{subsec:meeting}
To get a proof of concept for our \textit{SEnti-Analyzer} we tested the tool on a student software project meeting. The Software Engineering Group at Leibniz University Hannover yearly hosts a student software project for students in their last year of the bachelor in computer science. This way students can gain insight and experience in the professional software development process. Five to ten students work together on a software project, which are mostly applications for real life local customers (such as the \textit{Hannover Police Department} and the \textit{Hannover Medical School}). The whole project lasts one semester (approx. 15 weeks) with weekly meetings both team-internal and with the customer(s). The project team that participated in the case study worked on the \textit{VirtuHoS}-Project (Virtual House of Software), an application to virtually empathize the feeling of working together in an office with a decentralized development team. The team was tasked to create an editor for drawing a virtual office and creating the underlying semantic structure for further use by other groups using the Java programming language. Due to the ongoing Sars-CoV2 pandemic meetings could only be held virtually. For our case study, we recorded a 33-minute online meeting on 13th January 2021 in which all six team members participated. We collected written consent of each team member  allowing us to use the recorded audio files for research purposes and for scientific publications. The team participated voluntarily in the study and the participation had no influence on passing the course, on grades, etc.  

\subsection{Data Collection and Pre-Processing}

The meeting session was recorded digitally. We transcribed and classified the recordings by hand to get reasonable training data for the \textit{SEnti-Analyzer}. The team members used \textit{Discord} as their VoIP service and held the meeting together in a group call. A recording bot was used to record the meeting which enabled a multi-track recording separating each team member from another. At the end of the meeting, the team members were asked how they felt about the mood of the meeting (concerning the communication behavior). All team members agreed on the meeting communication being neutral to positive. A second prerecorded meeting from an older iteration of the student software project was also transcribed by hand increase the training set. The complete data set was then fed into the training function of the \textit{SEnti-Analyzer}, which finds new solutions by hyperparameter search for the included metrics extracted by natural language processing. For this training process, we used 1000 generations in total using an (1 + 1) evolutionary algorithm, thus only introducing one new population per generation and minimizing run-time.

\subsection{Data Analysis}
Both transcripts were split into single statements, which were then manually fitted with training labels to create the training data. A special training script loaded the whole data set into the training function of the \textit{SEnti-Analyzer} intending to learn a generalizing model. In total, our training data set consists of 712 manually transcribed and labeled statements, which follow the distribution shown in Table~\ref{table:distributions}. To validate our results, we used Fleiss' $\kappa$ as a statistical measure.

\section{Results}
\label{sec:results}

Based on the distribution of sentiment classes in the training set, an accuracy of 77.5\% would be possible by classifying each statement as neutral alone. Our model for the sentiment analysis tool provided by Klünder et al.~\cite{kluender2020identifying} however reached an accuracy of 81.8\% over the 712 statements from the training set, indicating a learning curve. The training of the model reached the peak fitness of 81.8\%  around the 800th to 900th generation. Out of a 10-minute audio file from the recorded meeting mentioned in subsection \ref{subsec:meeting} our tool extracted 140 statements. The distribution of the classified statements is shown in Table~\ref{table:distributions}.

\begin{table}[htbp]
\caption{Distributions of sentiment classes in training and test set}
\begin{center}
\begin{tabular}{|c||c|c|c|c|}
\hline
Statements in & \textbf{Total} & \textbf{Positive} & \textbf{Neutral} & \textbf{Negative}\\
\hline\hline
\textbf{Training set
} & 712 & 77 (10.8\%) & 552 (77.5\%) & 83 (11.7\%)\\
\hline
\textbf{Test set
} & 140 & 15 (10.7\%)& 124 (88.6\%)& 1 (0.7\%)\\
\hline
\end{tabular}
\label{table:distributions}
\end{center}
\end{table}

\subsection{Comparing Test Set Results to Training Data}
The trained model seemingly performs and generalizes well and the distribution of sentiment classes differs from that of the training data. Especially the relative distribution of the categories \textit{positive} and \textit{negative} changed widely from the training data. As Table~\ref{table:distributions} illustrates, the training set shows a virtually equal distribution of \textit{positive} and \textit{negative} statements (both within 11.2\% $\pm$ 0.5\% compared to the total training set size). The classified test set on the other hand shows a divergent distribution of the three sentiment classes, with the relative share of \textit{negative} statements decreased by 11\%. The \textit{neutral} class gains around 11.1\%, with only the \textit{positive} class staying at around the same percentage, only decreasing by 0.1\%. However, the classification of our tool directly corresponds to the feedback received by the team members, who told us they perceived the meeting communication as neutral to positive.

\subsection{Transcription Quality}
The speech recognition system we used showed difficulties when exposed to indistinct pronunciation or the high pace of speech of some of the team members. Usually, only single word errors occurred, but sometimes, when the pace was just too high for the speech engine or something said was obscure the transcript would differ so much that one could no longer deduce the actual statement from it. However, one has to consider that the German speech models we used had a given word error rate (WER) of 12.8\%. The English models offered directly by Mozilla on the DeepSpeech GitHub page are specified with a much lower WER of 5.97\% (release 0.8.2). Therefore much better transcription performance for English audio can be expected from the \textit{SEnti-Analyzer}.    

\subsection{SEnti-Analyzer Compared to Manual Classification}
To further verify the quality of our results we manually picked 50 from the 140 statements, which most matched the actual said and classified them again by hand to compare our classifications against the \textit{SEnti-Analyzer}. We did so according to our perception of the statement and the given context of previous statements, which the \textit{SEnti-Analyzer} does not yet consider. Table~\ref{table:results} compares the classifications taken by the \textit{SEnti-Analyzer} to the manual classifications. 

\begin{table}[htbp]
\caption{Comparison between classifications taken by the Senti-Analyzer and a human observer}
\begin{center}
\begin{tabular}{|c||c|c|c|}
\hline
Classification & \textbf{Positive} & \textbf{Neutral} & \textbf{Negative} \\
\hline\hline
\textbf{Software} & 10 (20\%) & 39 (78\%) & 1 (2\%) \\
\hline
\textbf{Manual} & 5 (10\%) & 45 (90\%) & 0 (0\%) \\
\hline
\end{tabular}
\label{table:results}
\end{center}
\end{table}

Remarkably, the classification of these statements by the \textit{SEnti-Analyzer} is more scattered than the classification by hand. Using Fleiss' $\kappa$, we calculated a $P_e$-value of $0.7282$ which means that the probability of a random match is 72.82\%. This can be traced back to the many matching classifications of both the \textit{SEnti-Analyzer} and the human observer in the category neutral, which reduces the value of $\kappa$. The total calculated $\kappa$-value is 0.56 and considered as an upper-moderate agreement according to the the scale provided by Landis and Koch~\cite{landis1977measurement}. A $\kappa$-value of 0.61 would already be considered as substantial agreement.  


\subsection{Summary}
Recapitulating, we fitted a model to our training data for the sentiment analysis tool provided by Klünder et al.~\cite{kluender2020identifying} to use in conjunction with our \textit{SEnti-Analyzer}. \textbf{The \textit{SEnti-Analyzer} automatically classified statements from an audio file in real-time producing results consistent with the feedback of the team members}.
We examined our results using the Fleiss' $\kappa$ measure and obtained an \textbf{upper-moderate agreement} as result.

\section{Interpretation}
\label{sec:interpretation}
We presented a concept which performs sentiment analysis on microphone audio in real time. We tested our concept on an exemplary tool and showed proof of concept on a real meeting from a student software project. This section discusses the findings with respect to the research question and threats to validity. In the end of this section, we point to future work.    

\subsection{Answering the Research Questions}
The findings and results we obtained can be used to answer the research questions we formulated at the beginning the following way:

\begin{itemize}[leftmargin=.38in]
    \item[RQ1:] We acquired proof of concept for our approach in the exemplary case study, and therefore successfully combined an automatic speech recognition system with our existing sentiment analysis tool~\cite{kluender2020identifying}. The developed tool is just exemplary and can be improved in many ways. Nevertheless, we are confident that this approach delivers valuable results and will lead to more productive working environments within software project teams in the future. 
    
    \item[RQ2:] According to Fleiss' $\kappa$, we reached a moderate agreement between the automatically produced results and the manual classification of a human observer. However, this outcome may be the result of numerous matching classifications in the \textit{neutral} class leading to a high $P_e$-value and thus a lower $\kappa$-value.
\end{itemize}

\subsection{Discussion}

Despite the preliminary nature of our study, we can observe two remarkable findings:

\begin{itemize}
    \item[(1)] Sentiment analysis can be applied to meetings, and
    \item[(2)] The application of sentiment analysis to verbal communication seems to be as meaningful as the application to textual communication.
\end{itemize}

The first finding appears to be a rather weak and not surprising observation, but offers a lot of potential for future work. Despite the fact that we do not have evidence supporting finding (2), this directly emerges from finding (1) with the fact that sentiment analysis itself has potential to support the collaboration in a team~\cite{obaidi2021development}. Therefore, it would be meaningful to start exploring the potential and to improve the polarity detection in meetings of software projects in a larger study. 

Nevertheless, besides the threats to validity which we discuss in the following section, there are four aspects that should be considered in the context of our study:

\subsubsection{High Chance of Random Agreement and Fleiss' $\kappa$}
The concept's applicability has been demonstrated by our exemplary tool, the \textit{SEnti-Analyzer}, alongside with the application on a real-world student software project meeting. The \textit{SEnti-Analyzer} provided feedback in line with the team's feedback after the meeting. For the chosen Fleiss' $\kappa$, a higher $P_e$-value (chance of a random match) leads to a lower $\kappa$-value. Unfortunately, due to the distribution of data in our test set, a lot of matching classifications in the class neutral occurred, thus leading to a high $P_e$-value of $0.7282$. The overall $\kappa$-value with 0.56 was, therefore, lower than expected concerning the overall agreement of 88\% (44 out of 50 total statements were classified identical both by the \textit{SEnti-Analyzer} and by hand). A more equally distributed test set would lead to a much lower $P_e$-value and thus facilitate a higher $\kappa$-value, and an even more meaningful result.

\subsubsection{Difficulty in Labeling Statements and Statement Context}
Klünder et al.~\cite{kluender2020identifying} already noted how labeling statements represents a difficult task, especially for a single human. Everybody has his/her own perception of the sentiment of a statement and, thus, two different people may choose a different sentiment class for the same statement. The training and test set which have been manually classified  for our research were only labeled by a single person. The size of the training data set was also limited by this factor. Furthermore, the statements were labeled concerning the context of the whole conversation, while the \textit{SEnti-Analyzer} (at the moment) only evaluates each statement on its own (without taking the context into account). Classifying the training and test sets by multiple persons choosing always the most voted class would reduce the impact of a single person's perception on the training data and results. The implementation of the concept of a statement context for the \textit{SEnti-Analyzer} is also imaginable.

\subsubsection{Domain Specificity}
Currently, the domain specificity of our tool is only given by the training data (all of the statements were taken from conversations out of software project meetings). For the German language, there are currently only general sentiment lexicons available, and no tools that would enable a domain-specific sentiment analysis. To improve the domain specificity and further tailor the tool to a software engineering context, the integration of a sentiment analysis tool designed specifically for software engineering would be beneficial. For the English language, one such example would be Senti4SD by Calefato et al.~\cite{calefato2018senti}.

\subsubsection{Impacts of the Sars-CoV2 Pandemic}
Because of the ongoing Sars-CoV2 pandemic, the meetings could only be held virtually which may have influenced our results. Some members lacked in audio quality and some had a notable level of background noise. Delays were also a problematic factor for the communication leading to multiple team members starting to talk at the same time or interrupting each other unintentionally. This would not happen that frequently in a real-world face-to-face meeting and a high-end audio setup in a conference room would provide better input audio quality enabling better results for the transcription of statements. The speech engine DeepSpeech is also constantly being improved, therefore future version upgrades may improve the transcription quality on their own.

\subsection{Threats to Validity}
Our case study results are limited to the used sample and cannot be generalized for other meetings. In this section, we summarize the most relevant threats to validity probably impacting our results.

We applied the \textit{SEnti-Analyzer} to a single meeting. The training set emerged from the manual labeling of two meetings. This small sample size was caused by the high effort of manually transcribing and labeling the meeting audio, and limits the statistical power of the results. In the same way, the used statistical measure, Fleiss' $\kappa$, may be unsuitable regarding the high number of matching classifications in the class \textit{neutral} by both observers (\textit{SEnti-Analyzer} and classification by hand). This aspect alone had a high influence on the calculated $\kappa$-value and the resulting strength of agreement. Because of the Sars-CoV2 pandemic, and the resulting curfew, the case study meeting had to be held online through VoIP software. Therefore, participants used their computers to attend the meeting. This influenced the experiment environment due to background noise, sounds from other rooms, noise from outside, and other static or interference noises. Delays over the VoIP also showed to be problematic while talking together, e.g., by cutting each other short unintentionally.

The student software project team recorded in the case study consisted of bachelor degree students in computer science. The prior knowledge about professional software development varied between team members. While some had already worked in private software corporations alongside their studies, others had programming knowledge only consisting of basic programming courses in computer science required to participate in the student software project. Therefore, the team members with less experience also did not use \textit{JIRA} or \textit{GitLab} prior to the software project. These variations may not influence the application of the \textit{SEnti-Analyzer}, but should be taken into account when interpreting the results. 

The \textit{SEnti-Analyzer} is currently not capable of differentiating voices, resulting in a transcription that consists merely of a concatenation of all recognized statements, instead of offering dialogue-like structuring. For the transcription to work as intended, it is necessary that only one person talks at a time, or otherwise,  the quality of the transcript will be compromised. The overall results are therefore limited by the currently free available transcription technology. Future research needs to focus on these issues.

\subsection{Future Work}
To reduce the possible impact of the threats to validity and to increase the reliability of our results, we propose the following steps for future work:

\begin{itemize}
    \item[(1)] Adjust the tool to English: As a first step, we want to adjust the tool to be applicable to meetings conducted in English, as both audio speech recognition and sentiment analysis tools provide better results in English than in German. This also helps extending the training set for the tool, as labeled data sets in English are way more frequently available.
    \item[(2)] Improve the reliability of the results: First and foremost, a (longitudinal) case study and a multi-case study are required to strengthen the results and to evaluate the usefulness of the application for the teams.
    \item[(3)] Taking facial and the tone expressions into account: As verbal communication is not the only communication used in meetings (rolling eyes or getting loud, e.g., also transport a lot of information), it would be interesting to also consider gestures, facial expressions, etc. 
    \item[(4)] Increase the granularity of the results, e.g., by distinguishing between different categories as proposed by Klünder et al. \cite{klunder2020affecting}. This helps pointing, for example, to destructive behavior which endangers project success by demotivating team members. 
\end{itemize}

\section{Conclusion}
\label{sec:conclusions}
Meetings represent a valuable way to communicate within development teams and are essential for every software project. To make software project meetings more effective and productive, and thus increase the overall mood and satisfaction of the project team, automated interaction analysis can be used. As the first step to our long-term research goal of an automated fine-grained interaction analysis, we introduce an approach combining prior interaction analysis research with the latest open source speech recognition achievements. The \textit{SEnti-Analyzer} processes meeting audio by cutting the conversation into single statements and transcribing them in real-time, before processing them using natural language processing.
The tool returns the classified statements and the overall meeting performance by showing the proportions of the sentiment classes \textit{positive}, \textit{negative}, and \textit{neutral}. This way, the project manager can gain additional informative feedback tracing the course of the meeting with little to no effort. Further actions can be taken due to the given resulting feedback, to improve future meeting behavior and communication. 

In a case study, we applied our tool to a real student software project meeting. The \textit{SEnti-Analyzer} delivered results that directly corresponded with the feedback the team members gave themselves. Using our results we could also verify moderate agreement of the classifications taken by the \textit{SEnti-Analyzer} in comparison to a human observer.

Overall, we propose to keep pursuing research on interaction analysis in software development teams using known sentiment analysis methods and machine learning algorithms to further expand the established concepts by integrating other components such as speech or gesture recognition. Automating tools for ease of use is also an important factor to disseminate interaction analysis in software development. 

\section*{Acknowledgment}
This research was funded by the Leibniz University Hannover as Leibniz Young Investigator Grant (Project \textit{ComContA}, 2020--2022).

\bibliographystyle{IEEEtran}
\bibliography{references}

\end{document}